\documentclass[twocolumn,prl,showpacs,superscriptaddress]{revtex4}
\usepackage[]{fontenc}
\usepackage[latin9]{inputenc}
\usepackage{bm}
\usepackage{amsmath}
\usepackage{graphicx}
\usepackage{amssymb}

\makeatletter

\usepackage{bm}
\usepackage{hyperref}

\@ifundefined{definecolor}
 {\usepackage{color}}{}

\usepackage{eepic}\usepackage{psfrag}
\usepackage{pstricks}%
\usepackage{pst-eps}%
\usepackage{pst-plot}%
\usepackage{pst-node}%
\usepackage{pst-poly}
\renewcommand{\d}{{\rm d}}
\renewcommand{\i}{{\rm i}}
\newcommand{\e}{{\rm e}}

\renewcommand{\Re}{{\rm Re}\;}
\renewcommand{\Im}{{\rm Im}\;}
\newcommand{\sgn}{{\rm sgn}}
\renewcommand{\arctan}{{\rm Tan}^{-1}}

\makeatother

\begin{document}

\title{Valley symmetry breaking in bilayer graphene: a test to the minimal
model }

\author{Masaaki Nakamura}

\affiliation{Department of Physics, Tokyo Institute of Technology, Tokyo 152-8551,
Japan}

\author{Eduardo V. Castro}

\affiliation{Instituto de Ciencia de Materiales de Madrid, CSIC, Cantoblanco,
E-28049 Madrid, Spain}

\affiliation{Centro de F$\acute{\imath}$sica do Porto, Rua do Campo Alegre 687,
P-4169-007 Porto, Portugal}

\author{Balázs Dóra}

\affiliation{Max-Planck-Institut für Physik Komplexer Systeme, Nöthnitzer Str.
38, 01187 Dresden, Germany}

\date{\today}
\begin{abstract}
Physical properties reflecting valley asymmetry of Landau levels in
a biased bilayer graphene under magnetic field are discussed. Within
the $4-$band continuum model with Hartree-corrected self-consistent
gap and finite damping factor we predict the appearance of anomalous
steps in quantized Hall conductivity due to the degeneracy lifting
of Landau levels. Moreover, the valley symmetry breaking effect appears
as a non-semiclassical de Haas-van Alphen effect where the reduction
of the oscillation period to half cannot be accounted for through
quasi-classical quantization of the orbits in reciprocal space, still
valley degenerate. 
\end{abstract}


\maketitle
Graphene, a two-dimensional hexagonal crystal of carbon atoms, has
attracted enormous attention in recent years \cite{scienceRev}. Its
quasiparticles are massless Dirac fermions propagating with a velocity
$1/300$ of the speed of light. This allows us to discuss and even
observe a variety of peculiar (ultrarelativistic) phenomena in a condensed
matter system, such as the anomalous quantum Hall effect (QHE), the
universal minimum conductivity, Klein tunneling, Zitterbewegung and
Schwinger's pair production \cite{review}.

Departures from strict two dimensionality promises interesting physics
as well. A bilayer graphene (BLG), which consists of a pair of single
graphene sheets bound by weak interlayer Van der Waals forces, has
also been studied extensively \cite{Novoselov-MMFKZJSG,McCann-F}.
Its low energy excitations possess chiral symmetry and a quadratic
spectrum, thus combining Dirac and Schrödinger like features. Moreover,
BLG offers interesting opportunities regarding device applications.
Breaking the layer symmetry opens an energy gap between valence and
conduction bands which in addition can be tuned by electric field
effect \cite{Ohta,McCann,Castro-NMPLNGGC1,Oostinga} --- a clear advantage
compared to current technology semiconductors. 

Establishing on a firm ground what is the minimal model describing
the system is an important step towards real device applications.
For instance, in the early experiments in BLG the gap was shown to
be tunable between zero (gapless) and mid-infrared energies \cite{Ohta,Castro-NMPLNGGC1},
which has been confirmed very recently by infrared spectroscopy \cite{ZhangNaturebbg,MakPRLbbg}.
The relevant minimal model has thus to correctly describe BLG up to
mid-infrared energies, which is a stringent restriction to the available
possibilities.

In the present paper we address the behavior of BLG in high magnetic
and electric fields applied perpendicularly. Using as a minimal description
the $4-$band continuum model with Hartree-corrected self-consistent
gap we predict new physics to take place at high enough fields. Specifically,
for magnetic fields between $\sim20-30\,\mbox{T}$ and electric fields
between $\sim0.5-1\,\mbox{Vnm}^{-1}$ (i.e. densities $\sim5-10\times10^{12}\mbox{cm}^{-2}$
in the standard $300\,\mbox{nm}$ $\mbox{SiO}_{2}$ back gate setup),
both within experimental reach, we have found that a significant valley
asymmetry is present with the following consequences: (i)~the period
of de Haas - van Alphen (dHvA) oscillations halves, in a clear non-semiclassical
behavior; (ii)~the QHE shows a new quantization rule for the Hall
conductivity given by $\sigma_{xy}=2\frac{e^{2}}{h}n$, with $n=0,\pm1,\pm2,\pm3\dots$
Observing experimentally this new behavior would put on a firm ground
our current understanding of BLG.


\emph{Model.}---We adopt the $4-$band continuum model as a key ingredient
in our minimal description. The Hamiltonian for one of the two valleys
($K$ point) is given by the following $4\times4$ matrix \cite{McCann-F,McCann,Nilsson}
whose elements correspond to the $A$ and $B$ sublattices of the
top and bottom layers ($A_{1},B_{1},A_{2},B_{2}$), \begin{equation}
\mathcal{H}_{K}=\left[\begin{array}{cccc}
\Delta & v\pi_{-} & 0 & t_{\perp}\\
v\pi_{+} & \Delta & 0 & 0\\
0 & 0 & -\Delta & v\pi_{-}\\
t_{\perp} & 0 & v\pi_{+} & -\Delta\end{array}\right],\label{Ham_bilayer.K}\end{equation}
where $\pi_{\pm}\equiv\pi_{x}\pm\i\pi_{y}$ and the vector $\bm{\pi}\equiv\bm{p}+e\bm{A}$
is the momentum operator in a magnetic field $\nabla\times\bm{A}=(0,0,B)$.
The parameter $v\simeq10^{6}\,\mbox{ms}^{-1}$ is the single layer
Fermi velocity and $t_{\perp}\simeq0.3\,\mbox{eV}$ the interlayer
hopping energy \cite{review}. Note that the zero entries in Eq.~\eqref{Ham_bilayer.K}
can be filled with next-nearest-interlayer hoppings originating trigonal
warping and electron-hole symmetry breaking effects \cite{Nilsson}.
At the energy scales we are interested in here, however, these terms
can be neglected in a minimal description \cite{McCann-F,Castro-NMPLNGGC2}.
The parameter $\Delta$ accounts for the layer asymmetry induced by
an external perpendicular electric field $\bm{E}=(0,0,\mathcal{E}_{{\rm ext}})$.
One way to relate $\Delta$ and the applied electric field $\mathcal{E}_{{\rm ext}}$
is by writing the local potential as a sum of two opposite contributions,
one coming from the electrostatic energy due to $\mathcal{E}_{{\rm ext}}$
which tries to polarize the system and a counteractive one originating
in the screening properties of the system. In the Hartree approximation
$\Delta$ can then be written as \cite{McCann,Castro-NMPLNGGC1,Min-S-B-M}
\begin{equation}
2\Delta=e\mathcal{E}_{{\rm ext}}c_{0}+\frac{e^{2}c_{0}}{2\varepsilon_{0}\varepsilon_{r}}\Delta n,\label{delta_vs_dn}\end{equation}
where $c_{0}\simeq3.4$~\AA{} is the interlayer distance, $\varepsilon_{0}$
is the permittivity of vacuum, $\varepsilon_{r}$ is the relative
permittivity of the system \cite{foot1}, and $\Delta n=n_{{\rm top}}-n_{{\rm bottom}}$
is the charge carrier imbalance between top and the bottom layers.
A self-consistent procedure is then followed since $\Delta n$ depends
directly on the weight of the wave functions in each layer. Instead
of $\mathcal{E}_{{\rm ext}}$ we use the density $n$ as an externally
tunable parameter. The two are related as $\mathcal{E}_{{\rm ext}}=en/(2\varepsilon_{0}\varepsilon_{r})$
in the standard back-gate configuration \cite{foot2}. 

\begin{figure}
\begin{centering}
\includegraphics[width=0.6\columnwidth]{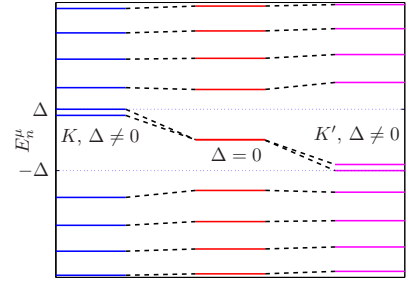} 
\par\end{centering}

\caption{(color online). Asymmetric structure of Landau levels at $K$ and
$K'$ points in bilayer graphene with energy difference $\Delta$.}

\label{fig:gapped_LL} 
\end{figure}

In zero magnetic field the system has a dispersion relation with the
characteristic double-minimum structure \cite{Guinea}. In the presence
of a finite magnetic field we can diagonalize the problem by going
to the Landau level (LL) basis \cite{Pereira-P-V}. Using the Landau
gauge $\bm{A}=(-yB,0,0)$, and noting that the commutation relation
between the momentum operators is $[\pi_{\pm},\pi_{\mp}]=\mp2eB\hbar$,
we define creation and annihilation operators of the harmonic oscillator
as $\pi_{\pm}\to\sqrt{2}\frac{\hbar}{l}a^{\dag}$ and $\pi_{\mp}\to\sqrt{2}\frac{\hbar}{l}a$
for $eB\gtrless0$, where $l^{2}\equiv\hbar/|eB|$ is the magnetic
length. Eigenvalues and eigenstates of Eq.~(\ref{Ham_bilayer.K})
are then obtained by assuming the wave function to be a linear combination
of the number states of the harmonic oscillator $|m\rangle$, with
integer $m\geq0$. The LLs , labeled by an integer $k\geq0$, are
given by \begin{equation}
E_{k}^{\mu}=\frac{\sqrt{2}\hbar v}{l}\lambda_{k}^{\mu},\qquad\mu=(s_{1},s_{2}),\label{eigenvalues}\end{equation}
with $r\equiv\frac{l}{\sqrt{2}\hbar v}t_{\perp}$ and $d\equiv\frac{l}{\sqrt{2}\hbar v}\Delta$.
The label $\mu=(s_{1},s_{2})$ specifies the outer and the inner bands
($s_{1}=\pm1$) and positive and negative ($s_{2}=\pm1$) energies,
respectively. For $k\geq2$ the parameter $\lambda$ is given by the
roots of the fourth-order polynomial \begin{multline}
\lambda^{4}-(2k-1+r^{2}+2d^{2})\lambda^{2}-2d\lambda\\
+[k(k-1)+d^{2}(d^{2}+r^{2}-2k+1)]=0,\label{eq_for_eigenvalue_a}\end{multline}
and the Landau states are \begin{equation}
|k,\mu\rangle\rangle=\left[\begin{array}{cccc}
\alpha_{k}^{\mu}|k-1\rangle, & \beta_{k}^{\mu}|k\rangle, & \gamma_{k}^{\mu}|k-2\rangle, & \delta_{k}^{\mu}|k-1\rangle\end{array}\right]^{T},\label{eigenstates_bi.1}\end{equation}
where the coefficients are given by $\beta_{k}^{\mu}=\frac{\sqrt{k}}{\lambda_{k}^{\mu}-d}\alpha_{k}^{\mu}$,
$\gamma_{k}^{\mu}=\frac{r\sqrt{k-1}}{(\lambda_{k}^{\mu}+d)^{2}-(k-1)}\alpha_{k}^{\mu}$,
$\delta_{k}^{\mu}=\frac{r(\lambda_{k}^{\mu}+d)}{(\lambda_{k}^{\mu}+d)^{2}-(k-1)}\alpha_{k}^{\mu}$
and the normalization condition. In addition there are the LLs $k=1$
and $k=0$. In the former case $\lambda$ is given by the roots of\begin{equation}
\lambda^{3}-d\lambda^{2}+(d^{2}+r^{2})(d-\lambda)=0,\label{eq:eigenvaluesLL1}\end{equation}
and the Landau state can be expressed as\begin{equation}
|1,\mu\rangle\rangle=\left[\begin{array}{cccc}
\alpha_{1}^{\mu}|0\rangle, & \beta_{1}^{\mu}|1\rangle, & 0, & \delta_{1}^{\mu}|0\rangle\end{array}\right]^{T}.\label{eq:eigenstate_bi.LL1}\end{equation}
Note that Eq.~\eqref{eq:eigenvaluesLL1} has only three roots, while
$\mu$ provides four labellings. We reserve the label $\mu=(-,+)$
for the $k=0$ LL, where $\lambda=+d$ provides the only eigenvalue.
We then write the respective Landau state as $|0,-,+\rangle\rangle\equiv\left[\begin{array}{cccc}
0, & |0\rangle, & 0, & 0\end{array}\right]^{T}$. The two states $|1,-,-\rangle\rangle$ and $|0,-,+\rangle\rangle$
are degenerate in the gapless case $d=0$ \cite{McCann-F}. For the
$K'$ point the Hamiltonian is given by Eq.~(\ref{Ham_bilayer.K})
with the replacement $\pi_{\pm}\to\pi_{\mp}$. Note that Eqs.~(\ref{eq_for_eigenvalue_a})
and~\eqref{eq:eigenvaluesLL1}, which determine the eigenvalues,
include a linear term in $d$ so that LLs for $K$ and $K'$ points
are not identical for $\Delta\neq0$ (see Fig.~\ref{fig:gapped_LL}).
The eigenvalues and eigenstates are then obtained by replacing $d\to-d$
in Eqs.~(\ref{eq_for_eigenvalue_a}) and~\eqref{eq:eigenvaluesLL1}
and by reversing the order of the elements in vectors~\eqref{eigenstates_bi.1}
and~\eqref{eq:eigenstate_bi.LL1}. As for the $k=0$ LL at $K'$
we reserve the label $\mu=(-,-)$ since the respective eigenvalue
has $\lambda=-d$, and write the eigenstate as $|0,-,-\rangle\rangle_{K'}\equiv\left[\begin{array}{cccc}
0, & 0, & |0\rangle, & 0\end{array}\right]^{T}$.

The final ingredient in the present model is the LL broadening. This
is taken into account through a constant imaginary retarded self-energy
$\Sigma_{{\rm ret}}=-i\Gamma$. Regarding the relevant value of $\Gamma$
we note that the dominant source of scattering originates from ripples
and charged impurities \cite{KG08}. Since BLG is less rippled than
its monolayer counterpart, and charged impurities do not penetrate
in between the layers, they mainly affect the layer closer to the
substrate, we take $\Gamma\simeq0.01t_{\perp}$ as a conservative
estimate. This phenomenological value is to be expected for a system
ballistic on the submicrometer scale and, moreover, when used to get
the longitudinal resistivity (by inverting the conductivity tensor
given below) it compares well with experiments in Ref.~\cite{Novoselov-MMFKZJSG}.


\emph{Self-consistent gap.}---The gap $\Delta$ defined in Eq.~\eqref{delta_vs_dn}
is obtained by writing the charge imbalance $\Delta n$ in terms of
the LL weight in different layers, \begin{align}
\Delta n= & \frac{1}{\pi l^{2}}\sum_{\mu}\sum_{k=0}^{\infty}\nonumber \\
 & \Bigl\{ f_{\Gamma}(E_{k}^{\mu})\bigl[(\alpha_{k}^{\mu})^{2}+(\beta_{k}^{\mu})^{2}-(\gamma_{k}^{\mu})^{2}-(\delta_{k}^{\mu})^{2}\bigr]\nonumber \\
 & -f_{\Gamma}(\bar{E}_{k}^{\mu})\bigl[(\bar{\alpha}_{k}^{\mu})^{2}+(\bar{\beta}_{k}^{\mu})^{2}-(\bar{\gamma}_{k}^{\mu})^{2}-(\bar{\delta}_{k}^{\mu})^{2}\bigr]\Bigr\},\label{dn_vs_wf}\end{align}
where the first and the second terms denote contributions from $K$
and $K'$ points, respectively. The factor $1/(\pi l^{2})$ is the
degeneracy of LLs per system volume. The function $f_{\Gamma}(E)$
is the Fermi distribution function in the presence of LL broadening,\begin{align}
f_{\Gamma}(x+\mu) & \simeq\frac{1}{2}-\frac{1}{\pi}\arctan\frac{x}{\Gamma}+\frac{\pi^{2}}{3}\frac{x\Gamma}{(x^{2}+\Gamma^{2})^{2}}(k_{{\rm B}}T)^{2},\label{eq:fGamma}\end{align}
valid in the low temperature limit $k_{{\rm B}}T\ll\Gamma$. In Eq.~\eqref{eq:fGamma}
the chemical potential $\mu$ is given by

\begin{equation}
n=\frac{1}{\pi l^{2}}\sum_{\mu}\sum_{k=0}^{\infty}\Bigl[\tilde{f}_{\Gamma}(E_{k}^{\mu})+\tilde{f}_{\Gamma}(\bar{E}_{k}^{\mu})\Bigr]\label{ng.1}\end{equation}
for a given density $n$, where $\tilde{f}_{\Gamma}(x)\equiv f_{\Gamma}(x)-f_{\Gamma}(x+\mu)$.
The solution is obtained numerically by solving Eqs.~\eqref{delta_vs_dn}
to~\eqref{ng.1} till self-consistency is achieved \cite{foot3}.

\begin{figure}
\begin{centering}
\includegraphics[clip,width=0.98\columnwidth]{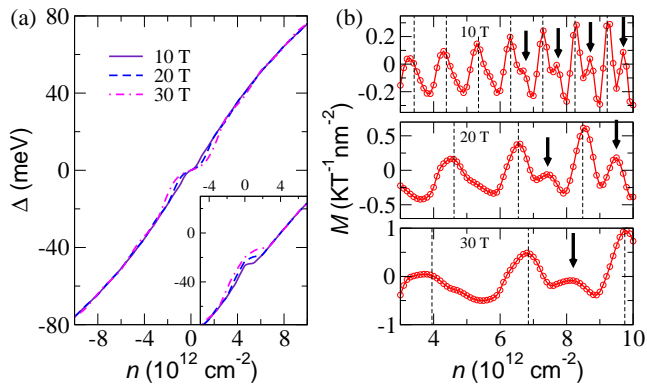} 
\par\end{centering}

\caption{(color online). (a) $\Delta$ vs $n$ for different values of magnetic
field. The inset shows the case of a fixed top-gate. (b)~Magnetization
vs $n$ for different values of magnetic field. The vertical lines
signal the semi-classical period.}

\label{fig:DeltadHvA} 
\end{figure}

In Fig.~\ref{fig:DeltadHvA}(a) we show the self-consistent asymmetry
gap $\Delta$ at $T=0$ as a function of electron density $n$. Increasing
the applied magnetic field increases screening around $n\approx0$,
which can be traced back to the valley (layer) asymmetry of the $0^{{\rm th}}$
and $1^{{\rm st}}$ LLs {[}see Eq.~\eqref{eq:eigenstate_bi.LL1}{]}.
In particular, the $0^{{\rm th}}$ LL appears fully polarized (layer
1 at $K$ and layer 2 at $K'$) thus giving the dominant contribution
to screening whenever the Fermi level lies in between the two. The
inset shows the case of a finite top-gate voltage fixed at $n_{{\rm tg}}=2\times10^{12}\,\text{cm}^{2}$
\cite{foot2}. The case of fixed filling factor and varying magnetic
field was considered in Ref.~\cite{Mucha} without LL broadening.

\emph{dHvA effect.}---As a consequence of LL quantization the magnetization
due to the orbital motion of electrons shows periodic oscillations
as a function of inverse magnetic field -- dHvA effect. Equivalently,
in systems where the carrier density can be tuned, similar oscillations
show up as $\mu$ changes. Here we address the dHvA effect in BLG
at fixed magnetic field $B$ and varying $n$ through a back-gate
voltage, taking into account that $\mathcal{E}_{{\rm ext}}$ changes
as the back-gate voltage is tuned. 

The magnetization for fixed density $n$ is given as $M=-\partial\Omega(B,\mu(n))/\partial B$,
where $\Omega$ is the Gibbs free energy and $\mu(n)$ is the chemical
potential satisfying Eq.~\eqref{ng.1} \cite{Sharapov-G-B}.  The
Gibbs free energy in the presence of LL broadening is obtained after
standard treatment and is given by \begin{equation}
\Omega(B,\mu)=\frac{\mathcal{V}}{4\pi^{2}l^{2}}\sum_{\nu}\sum_{k=0}^{\infty}\int_{-\infty}^{\infty}\d\varepsilon\e^{\varepsilon0^{+}}f(\varepsilon)g(\varepsilon-E_{k}^{\nu}+\mu),\label{free_ene_gamma}\end{equation}
where $f(x)\equiv(\e^{\beta x}+1)^{-1}$ is the Fermi distribution
function with $\beta^{-1}\equiv k_{\text{B}}T$ and $g(x)\equiv\pi-2\tan^{-1}(x/\Gamma)$,
the volume (area) of the system is $\mathcal{V}$, and $\nu$ includes
not only the band indices $(s_{1},s_{2})$ but also the valley index.
With the experimental conditions $\Gamma\gg k_{{\rm B}}T$ in mind
we consider the $T=0$ version of Eq.~(\ref{free_ene_gamma}), which
still has to be evaluated numerically along with the magnetization.
In particular, summation over the LLs is performed by introducing
a finite LL cut off $k_{c}$. We choose discrete values of the magnetic
field $B$ so that the number of states taken into account to obtain
$\Omega$ is preserved. Since $k>1$ LLs have the same degeneracy
proportional to the magnetic field, and since the degeneracy of $k=0$
and $k=1$ LLs add up to the same value, it is enough to choose the
magnetic field $B_{k_{c}}$ discrete values as $k_{{\rm c}}B_{k_{{\rm c}}}=\mathrm{Const}$
\cite{foot4}. A similar idea has been used for monolayer graphene
in Ref.~\cite{Koshino-A_2007a}. 

In Fig.~\ref{fig:DeltadHvA}(b) we show the dHvA effect as the carrier
density is changed at magnetic fields $B=10,20,30\,\mbox{T}$. The
vertical dashed lines signal the oscillation period given by the semiclassical
approach. The latter applied to a two-dimensional system in constant
magnetic field and varying density implies: a peak in magnetization
whenever the semiclassical orbit coincides with the Fermi surface;
a complete oscillation whenever the area enclosed by the Fermi surface
$A$ changes by the quantum of area enclosed by the semiclassical
orbit, $\Delta A=eB/h$ \cite{Onsager,Lifshitz-K}. Owing tho the
cylindrical symmetry of the bands {[}Eq.~\eqref{Ham_bilayer.K}{]}
the oscillation period can be expressed in terms of carrier density
as\begin{equation}
\delta n=2\frac{eB}{\hbar},\label{semiclassical}\end{equation}
where we used $A=|n|/(4\pi)$ and included spin and valley degeneracies.
When $n$ (i.e. the gap) is small the oscillations in magnetization
follow the semiclassical prediction {[}Fig.~\ref{fig:DeltadHvA}(b){]}.
With increasing $n$ (i.e. larger gap) we observe deviations from
the semiclassical value {[}arrows in Fig.~\ref{fig:DeltadHvA}(b){]}
and the new oscillation period is half of Eq.~\eqref{semiclassical}.
This provides a clear sign of the valley symmetry breaking, whose
experimental observation could be possible by diamagnetic current
measurements. Moreover, oscillations with a period which is half of
Eq.~\eqref{semiclassical} should also be observed in longitudinal
conductivity (Shubnikov-de Haas oscillations).

\begin{figure}
\begin{centering}
\includegraphics[clip,width=0.95\columnwidth]{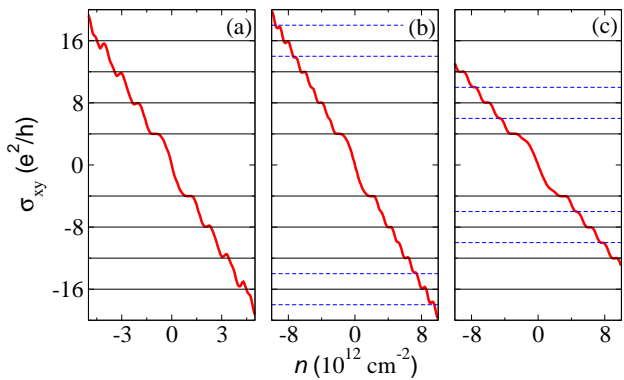}
\par\end{centering}

\caption{(color online). Hall conductivity vs electron density at different
magnetic fields: $B=10\,$T (a), $20\,$T (b), $30\,$T (c).}

\label{fig:sigma_xy} 
\end{figure}


\emph{QHE.}---The QHE is discussed by extending the results of Ref.~\cite{Nakamura-H-I}
for the gapless BLG case. We use the Kubo formula to get the longitudinal
and Hall conductivities, \begin{equation}
\Re\sigma_{ij}(\omega)=\frac{\Im\tilde{\Pi}_{ij}(\omega+\i\eta)}{\hbar\omega},\label{Kubo_form}\end{equation}
where $\tilde{\Pi}_{ij}(\omega)\equiv\Pi_{ij}(\omega)-\Pi_{ij}(0)$
with $\{i,j\}\in\{x,y\}$. The polarization function $\Pi_{ij}(\omega)$
is given by the current-current correlation function, and obtained
as the analytical continuation of the Matsubara form: \begin{align}
\lefteqn{\tilde{\Pi}_{ij}(\i\nu_{m})=}\label{hall_conductivity_xx_derivation2}\\
 & -\frac{e^{2}}{2\pi l^{2}\beta\hbar}\sum_{n=-\infty}^{\infty}\sum_{k,l}\sum_{\mu,\nu}\frac{\langle\langle k,\mu|\gamma_{i}|l,\nu\rangle\rangle\langle\langle l,\nu|\gamma_{j}|k,\mu\rangle\rangle}{(\i\tilde{\omega}_{n}-\tilde{E}_{k}^{\mu})(\i\tilde{\omega}_{n}^{+}-\tilde{E}_{l}^{\nu})},\nonumber \end{align}
where $\mu$ and $\nu$ include not only the band indices but also
the valley index. The matrix $\gamma_{i}$ is defined by $\bm{\gamma}\equiv\nabla_{\bm{p}}{\cal H}$,
and $\hbar\tilde{E}_{k}^{\mu}\equiv E_{k}^{\mu}$ with LL $E_{k}^{\mu}$
as in Eq.~\eqref{eigenvalues}. We have defined $\i\tilde{\omega}_{n}\equiv\i\omega_{n}+[\mu+\i\,\sgn(\omega_{n})\Gamma]/\hbar$
and $\omega_{n}^{+}\equiv\omega_{n}+\nu_{m}$ with $\omega_{n}$ ($\nu_{m}$)
as the Matsubara frequency of fermions (bosons). 

Numerical results for the Hall conductivity as a function of carrier
density (no top gate is assumed) are shown in Fig.~\ref{fig:sigma_xy}
for different magnetic fields \cite{foot5}. For $B=10\,\mbox{T}$
the Hall conductivity follows the gapless BLG quantization rule $\sigma_{xy}=4ne^{2}/h$
with $n=\pm1,\pm2,\pm3\dots$ \cite{Novoselov-MMFKZJSG}. For $B=20-30\,\mbox{T}$
new quantized Hall steps {[}dashed horizontal lines in Figs.~\ref{fig:sigma_xy}(b)
and~\ref{fig:sigma_xy}(c){]} appear at high carrier densities (large
asymmetry gap). The new quantization rule reads $\sigma_{xy}=2ne^{2}/h$
and is a direct consequence of the degeneracy lifting between LLs
from $K$ and $K'$. Including a top gate enables the observation
of the new quantization rule near and at the neutrality point \cite{Tutuc}.


\emph{Conclusion.}---A remarkable property of BLG is the possibility
to open and tune an energy gap by breaking the layer symmetry. At
finite magnetic fields this layer asymmetry translates into an asymmetry
between the two valleys, no longer protected by time-reversal symmetry.
Using the 4-bands continuum model with self-consistent gap and Landau-level
broadening we have shown that the effects of valley symmetry breaking
are manifested in non-semiclassical oscillations of the magnetization
and anomalous quantized Hall steps. Their experimental detections
is within reach, and would allow for a critical test of the minimal
model for BLG.


We thank M. Koshino and M. A. H. Vozmediano for illuminating discussions.
MN is supported by Global Center of Excellence Program \char`\"{}Nanoscience
and Quantum Physics\char`\"{}. EVC is financially supported by the
Juan de la Cierva Program (MCI, Spain). BD was supported by the Hungarian
Scientific Research Fund under grant number K72613 and by the Bolyai
program of the Hungarian Academy of Sciences. MN and EVC acknowledge
the visitors program at the Max-Planck-Institut f\"{u}r Physik komplexer Systeme,
Dresden, Germany.



\begin{thebibliography}{99}
\bibitem{foot1}
The value of $\varepsilon_r$ depends on the substrate. For a SiO$_2$/air interface 
$\varepsilon_r \approx 2.5$. Here we use $\varepsilon_r \approx 1$ as a reference.
\bibitem{foot2}
In dual-gate cases $\mathcal{E}_{\rm ext}$ and $n$ are independent: $\mathcal{E}_{\rm ext} = e(n_{\rm bg} - n_{\rm tg})/(2\varepsilon_0 \varepsilon_r)$ and $n=n_{\rm bg} + n_{\rm tg}$, where $n_{\rm bg}$ and $n_{\rm tg}$ are back- and top-gate induced densities \cite{Oostinga,Tutuc}.
\bibitem{foot3}
The LL sum in Eqs.~\eqref{dn_vs_wf} and~\eqref{ng.1} is cutt off at $k_c \approx 1000$. A similar approach was followed in Ref.~\cite{Mucha}.
\bibitem{foot4}
We use $k_c \approx 1000$ and checked consistency of results.
\bibitem{foot5}
Summing over 100 LLs is enough for the conductivity.
\end{thebibliography}
\end{document}